\documentclass[notoc]{JHEP3}
\usepackage{amssymb, epsfig}

\newcommand{\newc}{\newcommand}
\newc{\beq}{\begin{equation}}
\newc{\eeq}{\end{equation}}
\newc{\beqa}{\begin{eqnarray}}
\newc{\eeqa}{\end{eqnarray}}
\newc{\IM}{\mbox{\sl{Im}}}
\newc{\RE}{\mbox{\sl{Re}}}
\newc{\nonr}{\nonumber}
\newc{\hs}{\hskip 3mm}
\newc{\ra}{\rightarrow}
\newc{\TR}{\mbox{\sl{Tr}}}
\newc{\tri}{\triangle}
\newc{\szorb}{$S_1/(Z_2\hskip -1mm \times \hskip -1mm Z_2^{\prime})$ }

\title{
 Neutrino Masses in 5D Orbifold $SU(5)$
 Unification Models without Right-handed Singlets }
\author{We-Fu Chang and John N. Ng \\
TRIUMF Theory Group, 4004 Wesbrook Mall, Vancouver, B.C. V6T 2A3 CA \\
E-mail:\email{wfchang@triumf.ca, misery@triumf.ca} }

\abstract{
 We explore a mechanism for radiatively generating
neutrino Majorana masses in a 5 dimensional  orbifold $SU(5)$ unification
model without introducing right-handed singlets. The model is
non-supersymmetric and the extra dimension is compactified via  a
\szorb  orbifold geometry.
The necessary lepton number violating interaction arises from the
Yukawa interactions either between a 10-plet or a 15-plet bulk scalar
field  and the fermion quintuplets which are residents on the $SU(5)$ symmetrical
brane located at one of the orbifold fixed points.
The model is engineered to give
realistic charged fermion masses and mixing and in the same time
avoiding the rapid proton and neutron decays by geometric construction.
The gauge unification can be maintained  by adding extra fermion
or scalar fields. The unification scale is found to be larger
then $10^{15}$ GeV by adding a bulk vector decuplet pair whose zero mode has
masses  around $10\sim 100$ TeV range.
We found that neutrino mass matrix of the normal hierarchy type is
favored by using 15-plet scalar. We give a solution of this type which has
detectable $\mu\ra 3e$ transition. On the other hand, by introducing 10-plet
scalar, the leading neutrino mass matrix can only be inverted
hierarchical and gives at most bi-maximal mixing.}


\begin{document}
\section{Introduction}
Recent measurements of atmospheric and solar neutrino fluxes at the
Super Kamiokande \cite{SuperK} and the Sudbury Neutrino Observatory \cite{SNO}
have provided compelling evidence for neutrino masses and neutrino oscillations.
This received further support from the reactor KamLand experiment \cite{KamL}.
Furthermore, the Wilkinson Microwave Anisotropy Probe \cite{WMAP} imposes the constraint
that the sum of neutrino masses to be less than $.75$ eV.
Such a small value for neutrino masses is  generally considered to be a harbinger of
new physics beyond the standard model (SM) and the existence of a new scale between
the Fermi and the Planck scale.
In particular if the three neutrinos involved in weak interactions are Majorana
in nature then clearly new physics is at play.
The most popular suggestion of generating neutrino masses in the milli-electronvolt range
is grand unified theories (GUTs) via the seesaw mechanism with or without supersymmetry.
Central to this idea is the introduction of one right-handed singlet neutrino per
family of the SM fermions with a mass near the GUT scale.
This is natural in $SO(10)$ models since its fundamental $\mathbf {16}$ representation
encompasses this singlet with the 15 fermions of the SM.
For a recent review of neutrino masses in grand unified models see \cite{Alta03}.
On the other hand small Dirac neutrino masses is considered unnatural due to the extreme
fine tuning required. However, in theories with extra dimensions this
can be generated   by allowing the singlet neutrinos to be bulk fields.
A small Yukawa coupling can be obtained due to the volume dilution factor if the
extra dimensions are  sufficiently large \cite{Dienes:1999}.
In both cases right-handed singlet fields $N_R$ are necessary.

In this paper we study the construction of  neutrino mass  without
the benefit of $N_R$ in the context of grand unified $SU(5)$ models with the
minimal particle content\footnote{ It is also possible to generate neutrino masses
without using  $N_R$ by R-parity violating interaction in minimal supersymmetric
standard model\cite{RPX}. Here we concentrate on non-supersymmetric models and
leave questions such as quantum stability of scalar fields and naturalness issues
as unsolved. }.
This is a fundamentally different mechanism from the above mentioned constructions.
Since the neutrinos in this scenario are Weyl particles of the SM and the resulting
neutrino mass matrix is necessarily Majorana.
In conventional four dimensional (4D) field theories one can use SM Higgs triplet to
achieve this.
However, the vacuum expectation value (VEV)  of the triplet must be fine tuned to
small values in order not to upset the highly  successful custodial $SU(2)$ relation
of the SM gauge bosons as well as to generate of a sufficiently small neutrino mass.
A second method is to radiatively generate neutrino masses  at 1-loop.
The prototype model was constructed some time ago  \cite{Zee} and the
crucial ingredient is the introduction of a $SU(2)$ singlet scalar field with non trivial
weak hypercharge.
The original version  of the model gives a $3\times 3$  neutrino mass matrix
with zero diagonal elements and thus leading to bi-maximal neutrino mixings \cite{Zeeph}.
This is ruled out by the data. However, simple phenomenological modifications
can bring it to agree with observations \cite{He}.
All these constructions suffer from being rather ad hoc.
It will be interesting if one can incorporate these attempts into the theoretically
well motivated unification models. A more modern formulation of this makes use of
progress in recent works on extra dimensions and the
brane world scenario.
It has been shown that the technique of orbifold projections applied to
GUTs \cite{kawamura} can solve some of the long standing problems such as
doublet-triplet splitting that plagued 4D $SU(5)$.
This is further applied  to the flavor problem by various workers \cite{Flavorb}.
However, in these works  neutrino mass matrices are constructed using
right-handed singlets along the line of the seesaw mechanism.

In a previous paper \cite{Chang:2003sx} we constructed a viable model of radiative
neutrino masses with minimal SM matter in a five dimensional (5D) field theory
on the orbifold \szorb with bulk $SU(3)_W$ gauge symmetry.
The $SU(3)_W$ unifies the $SU(2)\times U(1)$ electroweak symmetry of the SM \cite{SU3W}.
The crucial observation is that the SM lepton doublet and charged lepton
singlet can naturally be embedded in the fundamental representation of $SU(3)_W$.
With Higgs fields in the ${\mathbf 6}$ and ${\mathbf 3}$ representations lepton number
violating interactions can be constructed.
Tree level masses are forbidden by orbifold projections and thus
avoiding the fining tuning of VEV for small neutrino masses.
Neutrino masses can be generated at  1-loop level.
The scale of neutrino masses is small compared to the weak scale due to three factors:
(i)  the loop factor, (ii) the inverse of the compactification radius, $R$, which
controls the volume dilution factor and (iii) small Yukawa couplings of the charged leptons.
We found neutrino masses of order $0.01$ eV without much fine tuning.
Interestingly the solutions that satisfy the observed mixing parameters are found for the
inverted mass hierarchy type if Yukawa couplings are not  fine tuned.

We continue this  investigation for the GUT theory of $SU(5)$ which is theoretically
well motivated.
This poses new challenges  since the fundamental representation ${\mathbf 5}$ unifies
the lepton doublet $(\nu \; e)_L$ with the right-handed down type quark $d_R$.
We found two options for the Higgs fields that can give rise to lepton number violating
interactions essential for our mechanism.
Strictly speaking the concept of lepton number is not fundamental in these theories;
however, we find it convenient to use it both for guiding model construction as
well as in  navigating the tight constraints imposed by the many rare decay experiments.
The Higgs fields are the $\mathbf{15}$ or  $\mathbf{10}$ representation.
Another problem we encountered in this construction is to maintain gauge coupling
unification since it is well known that exotic Higgs contributes to the running of the
coupling constants.
This solved by introducing bulk fermion fields or additional Higgs fields.

This paper is organized as follows: In section 2,
we will review the setup of 5D orbifold $SU(5)$ theory.
The details of the model are given here.
The construction of the Majorana neutrino masses through the Yukawa interaction of bulk
$\mathbf{10}$ or $\mathbf{15}$ Higgs is made explicit in section 3.
The gauge unification question will be addressed in section 4.
The rich  exotic Higgs sector leads to interesting phenomenology and will be
discussed in section 5.
Finally we give our conclusions in section 6.

\section{5D Orbifold $SU(5)$ Model}

We begin by a brief review of the 5D $SU(5)$ unification model defined on the orbifold
\szorb \cite{SU5orb} with coordinates $x^{\mu} (\mu =0,1,2,3)$ and the extra spatial
dimension which is denoted by $y$.
The circle  $S_1$ of  radius $R$, or $y=[-\pi R, \pi R]$, is orbifolded by parity
$P: y\leftrightarrow -y$ transformation.
The resulting space is divided by a second $Z_2^{\prime}$ acting  on
$y'= y - \pi R/2$  as $P': y'\leftrightarrow -y'$ to give the final geometry.
On this orbifold, the Fourier decomposition is summarized in Table 1.

\TABULAR[htb]{c c c}{
  \hline
  $(P,P')$ &  form & mass  \\
  \hline \hline
  $(++)$ & $\frac{\sqrt2}{\sqrt{\pi R}}[A_0(x)+\sqrt{2}\sum_{n=1} A^{++}_{2n}(x) \cos\frac{2ny}{R}] $
   & $ 2n/ R$  \\
  $(+-)$ & $\frac{\sqrt2}{\sqrt{\pi R}}[\sqrt{2}\sum_{n=1} A^{+-}_{2n-1}(x) \cos\frac{(2n-1)y}{R}] $
   & $(2n-1)/ R$  \\
  $(-+)$ & $\frac{\sqrt2}{\sqrt{\pi R}}[\sqrt{2}\sum_{n=1} A^{-+}_{2n-1}(x) \sin\frac{(2n-1)y}{R}] $
  & $(2n-1)/ R$  \\
  $(--)$ & $\frac{\sqrt2}{\sqrt{\pi R}}[\sqrt{2}\sum_{n=1} A^{--}_{2n}(x) \sin\frac{2ny}{R}] $
   & $2n/R$  \\ \hline }
{ KK decomposition of a bulk field $A(x,y)$ with parities $(P,P')$.}

Under the  $Z_2 \times Z_2^{\prime}$ transformation, there are two fixed points at
$y=0$ and $y=\pi R/2$ denoted by $y_S$ and $y_G$ respectively.
The following two parities matrices are chosen for the parity transformations:
\beqa
P&=&\mbox{diag}\{+++++\},\nonr\\
P'&=&\mbox{diag}\{---++\}.
\eeqa
These determine the Kaluza-Klein decompositions of a generic bulk field $A(x,y)$ (see Table 1).
The parity assignments of a bulk field are  chosen by phenomenological considerations.
Indeed for a given multiplet different components can have different  parities
under $Z_2 \times Z_2^{\prime}$.
In particular, for the $SU(5)$ gauge fields $A_M (M=\mu, y)$, the following parities
are used
\beqa
Z_2&:& A_\mu\ra + P A_\mu P^{-1},\hs  A_y \ra -P A_y P^{-1} \nonr\\
Z'_2 &:& A_\mu\ra + P' A_\mu P^{'-1},\hs A_y \ra \pm P' A_y P^{'-1}.
\eeqa
where we have suppressed the group index.
Since  the fifth components  vanish at the $y_S$ fixed point
they  will not enter the low energy effective theory.
With this parity assignment,  when decomposed into the SM subgroup
$G=SU(3)\times SU(2)\times U(1)$, the $SU(5)$ gauge bosons have following components:
\beq
24=(8,1,0)_{++} + (1,3,0)_{++} + (1,1,0)_{++}+
\left( 3, 2, -\frac56\right)_{+-} + \left(\bar{3}, 2, \frac56\right)_{+-}
\eeq
where hypercharge is normalized to $Y=Q-T_3$.
The parities $(P,P')$ are shown  as subscripts. The first three terms are the SM gauge
bosons. With the  assigned  $(++)$ parities  they will  have zero modes.
On the other hand, the last two terms represent the gauge boson $X$s and $Y$s
which have no zero modes since their parities are $(+-)$.
They are KK excitations.
At the symmetry broken brane, $y_G$, even these Kaluza-Klein(KK) excitations
 vanish. So at $y_G$, the symmetry is always $G$.
On the other hand, at the symmetric brane, $y_S$, $SU(5)$ is still a good
symmetry. At the low energy, only  the zero modes of $G$ can
be observed. When energy is higher then $1/R$, the appearance of KK
excitations of $SU(5)/G$ plus the KK excitations of $G$ will
restore the full $SU(5)$ symmetry.
By geometrical construction, the gauge $X,Y$ gauge bosons are massive since they are
KK excitations and the lightest ones have masses  $\sim 1/R$. At $y_G$ the bulk $SU(5)$
symmetry is broken by the orbifold boundary conditions.
So there is no need to  introduce the $\mathbf{24}$ as in the 4D case.
Thus, the triplet-doublet problem \cite{DT-splitting} is solved naturally \cite{kawamura}.
To break $G$ to $SU(3)\times U(1)$  and to generate fermions masses, one still needs
one Higgs in $SU(5)$ fundamental representation.

The SM fermions can be grouped into the standard $\mathbf{\bar{5}}$ and $\mathbf{10}$
( henceforth denoted as $F$ and $T$ respectively ) representations:
$F_i=(D^c_i, L_i)$ and $T_i=(U^c_i, Q_i, E^c_i)$
where we use left-handed chiral fields with obvious notations and $i=1,2,3$ is the
family index whereas the superscript $c$ denotes charge conjugation.
In a 5D theory they can either be bulk fields or brane fields.
If the matter fields  are placed at the symmetric brane they will enjoy the
merits and suffer the drawbacks of $SU(5)$ GUT: the quantization of hypercharge,
the unification of gauge couplings, the prediction of mass ratio
of down-type quark and charged lepton, and the baryon number violating decays
\cite{GG}.
In a  5D model the possibility that some fermions can be bulk fields and
plus the existence of different fixed points to locate fermions can be used to retain
the successful mass relationships and avoid some difficulties such as the proton decay problem.
For simplicity we shall assume that the $F_i$ are localized on a brane at $y_S$.
Hence, there will be no KK excitations of neutrinos.

Because the prediction of $m_b/m_\tau$ agrees quite well with experiment it is reasonable to localize
 both $T_3$ and $F_3$ on the brane at $y_S$.
On the other hand, we can make  $T_1$ a bulk  field so as to avoid rapid proton decay,
see Fig.\ref{fig:protondecay}.
This is so  because  both $T_1$ and the  $X, Y$ bosons are bulk fields and their
interaction must honor the conservation of KK number; i.e.
the  $X, Y$ bosons are KK modes they do not couple to two up-quark zero modes.
Therefore, all the sub-diagrams vanish in the 5D model.
Similar  diagrams which lead to baryon number violating  neutron decays are also forbidden
by the KK number conservation and hence matter stability is not violated.

\FIGURE[ht]{
\epsfxsize=2.6in
\centerline{\epsfbox{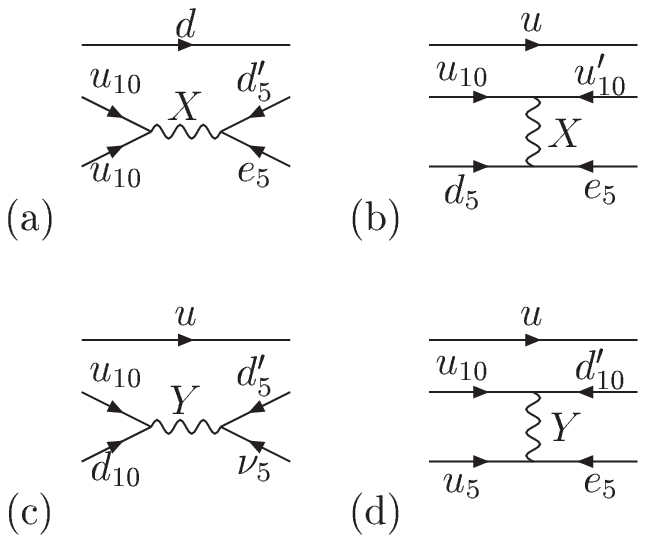}}
\label{fig:protondecay}
\caption{The proton decays through exchanging
the $X$ and $Y$ gauge bosons. The subscripts denote $SU(5)$ representations.}}

As pointed out in \cite{Hall:2001xb}   the parities of bulk fermion fields
pose further complications. If  $T_1$ has parities  $(++)$, then
the parities of its components will be $(U^c_1(++),Q_1(+-),E^c_1(++))$.
Clearly the quark doublets will not have zero modes and hence not acceptable.
To complete the SM particle spectrum another $T'_1$ of parity $(+-)$ to be
introduced such that the combination of the zero modes of $(U^c_1,Q'_1,E^c_1)$
and brane field $F_1$ constitutes the first SM family.
In doing so the $SU(5)$ mass relation will not be obeyed which is good phenomenologically.
Henceforth we shall adopt the proposal that both $T_1$ and $T_2$ are bulk fields and
no $SU(5)$ mass relations holds for the first two generations.

Letting $T_1$ and $T_2$ go into bulk immediately implies a hierarchy between
4D effective mass of the third and the first two generations.
The  5D Yukawa interaction can be written as
\beq
{\cal L}_5 \supset
{\tilde{f}_{u ij}\over \sqrt{2M^*}}\overline{\psi}^{\mathbf{c}}_{10i} \psi_{10 j}H_5
+ {\tilde{f}_{d ij}\over \sqrt{2M^*}}\overline{\psi}^{\mathbf{c}}_{\bar{5}i} \psi_{10 j}H_5^*
+ H.c.
\label{eq:5DYuk}
\eeq
with group indices suppressed and $M^*$ denotes the fundamental scale.
Because $T_1$ and $T_2$ live in the bulk, by naive dimension analysis,
the 4D effective Yukawa coupling  for  down-type quarks naturally exhibit
the following hierarchy patterns
\beq
\label{eq:masspattern-D}
y_d \sim c_d \left(\begin{array}{ccc}
  \epsilon & \epsilon & 1\\
  \epsilon & \epsilon & 1\\
  \epsilon & \epsilon & 1\\
\end{array}\right)
\eeq
where the volume dilution factor $\epsilon\sim (M_c/M_*)^{1/2}\sim 0.1$
and $c_d$ is a common factor.
For the up-quarks, we adopt the following structure for 4D effective
Yukawa couplings as proposed by \cite{Hall:2001xb}:
\beq
\label{eq:masspattern-U}
y_u \sim c_u \left(\begin{array}{ccc}
  \epsilon^2 & \epsilon^2 & \epsilon\\
  \epsilon^2 & \epsilon^2 & \epsilon\\
  \epsilon & \epsilon & 1\\
\end{array}\right).
\eeq
The volume dilution mechanism  still works for the Yukawa
interaction between  one brane and one bulk
fermions, i.e. $y_{u13}, y_{u23}, y_{u31}$ and $y_{u32}$.
But one needs to assume the Yukawa couplings between $T_1, T_2$  to be
two order of magnitude smaller then the others in order to have the structure
of Eq.(\ref{eq:masspattern-U}).
Given the totally different nature of $T_1, T_2$ from other
fermions, this assumption is an ad hoc one and we have no deeper understanding
why this has to be so.

The 5D bulk lagrangian is  $Z_2 \times  Z_2'$ symmetric.
The question now arises: what parities should  be assigned to the
brane fermions?
In our case  the brane fermions will reside at $y=0$, or
equivalently $y=\pi R$, brane. They have definite parities
under $Z'_2$ transformation, such as
\beqa
Z_2 &:& \Psi_{\bar{5}} \ra + P\Psi_{\bar{5}},\hs
      \Psi_{10}\ra + P^T \Psi_{10} P\\
Z_2' &:& \Psi_{\bar{5}} \ra \pm P'\Psi_{\bar{5}},\hs
      \Psi_{10}\ra \pm P'^T \Psi_{10} P'
\eeqa
So there are two ways of how the $Z'_2$ parities  can
be assigned to brane fermions:
(1) $(T,F)=(Q,U,E, D,L)=\pm(+--,-+)$
and (2) $(T,F)=(Q,U,E, D,L)=\pm(+--,+-)$.
In order that  the 5D lagrangian shall respect the symmetry of the theory
we can simply make the follow manipulation on the brane term:
\beq
\int d^4x \int dy \frac12 \{\delta(y) \pm \delta(y-\pi R)\}  L_{\pm},
\eeq
where subscript $+$ and $-$ stand for $Z'_2$-even and -odd
respectively. Since our effective physical space is restrict
to $y \in [0,\pi R/2]$, the choice of  parity for  the brane fermions do not make
any difference.

Now we turn to the discussion of electroweak symmetry breaking.
This is done by the  bulk scalar in the $\mathbf{5}$
representation, $H_5$, to reduce $G$ to $SU(3) \times U(1)_Q$ and to
give charged fermion masses as seen in Eq.(\ref{eq:5DYuk}).
However,  $H_5$ contains the color Higgs $H_c$ with quantum number
$(3,1,-1/3)$ and the ordinary SM Higgs doublet $H_w$. The  $(+-)$ parity
of $H_c$  forbids the existence of zero mode and hence no
light color scalars on the $y_S$ brane. Denoting  $H_w=(h^+,
h^0/\sqrt{2})^T$ as in SM, the Higgs doublet develops VEV
$\langle H^T_w \rangle = (0, v_b^{3/2}/\sqrt{2})$
which breaks the electroweak symmetry as usual.
Integrating out the fifth dimension, the effective gauge coupling can be
used for  relating  the 5D and 4D gauge couplings,
 $g_4= \tilde{g} /\sqrt{\pi R M^*}$.
The resulting $W$ boson mass is found to be $M_W^2= g_2^2 \pi R v_b^3 /2$,
 or $ (2\pi R v_b^3)^{1/2} =v_0 \sim 250$ GeV.
With the above parameter substitution, we arrived at
the effective 4D Yukawa interaction relevant to  charged fermions masses:
\beqa
{\cal L}_Y^4 &=&y_{u,ij} {v_0 \over \sqrt{2}} \bar{u'}_{R i} u'_{L j}
+ y_{d,ij} {v_0 \over \sqrt{2}}( \bar{d'}_{R i} d'_{L j}+ \bar{e'}_{R j} e'_{L i})
 + H.c.\\
&=& \bar{u'}_{R} {\cal M}_{u} u'_{L}
 +\bar{d'}_{R} {\cal M}_{d} d'_{L}
  + \bar{e'}_{R} {\cal M}_{e} e'_{L}
  + H.c.
\eeqa
Where the mass matrices can be identified as:
\beq
{\cal M}_u=  {v_0 \over \sqrt{2}} y_{u},\:
{\cal M}_d=  {\cal M}^T_e=  {v_0 \over \sqrt{2}} y_{d}.
\eeq

The mass matrices are diagonalized by bi-unitary rotation:
\beqa
U_R {\cal M}_u  U_L^\dag = {\cal M}_u^{dig},\hs
V_R {\cal M}_d V_L^\dag  = {\cal M}_d^{dig},\\
V_L^* {\cal M}_e V_R^T= {\cal M}_e^{dig},\hs
V_{CKM}= U_L^\dag V_L
\eeqa
and  the leptons can be rotated into their  mass eigenstates by
\beq
L_i= (V_R^*)_{ij} L'_j, \hs e_{R i}= (V_L^*)_{ij} e'_{Rj}.
\eeq
Note that the neutrino is massless at this point, so we have
the freedom to apply left-handed rotation  to the entire lepton $SU(2)$
doublet and make the charged current interaction diagonal.
Also the Eq.(\ref{eq:masspattern-D},\ref{eq:masspattern-U}) can not be taken literally
otherwise the first two generations are degenerated.
It shall be understood that every entity exhibits a small correction to the leading
term exhibited.
Numerical check shows that by allowing a $10\%$ correction  of each entry,
it's easy to get the resulting mass hierarchy:
\beqa
m_b : m_s :m_d = m_\tau :m_\mu: m_e \sim 1:\epsilon
:\epsilon^2,\\
m_t : m_c : m_u \sim 1:\epsilon^2:\epsilon^4,
(V_{us}, V_{cb}, V_{ub})\sim (\epsilon, \epsilon, \epsilon^2).
\eeqa
In our numerical experiments about one third of the statistical samples provide CKM
matrix very close to experimental values. Hence,  these two general
mass matrix patterns successfully yield the observed charged
fermions mass hierarchy and CKM mixing angles.

Since the lepton mixing is crucial in the mechanism for
generating  neutrino mass  which to be discussed in section 3, more
insight of the mixing matrix is required.
It's easy to see that $V_L\sim diag(1,1,1)$, so the CKM is mainly controlled
by the mixing of up-quark, $U_L$.
The $V_R$ is to diagonalize the almost uniform mass square matrix
\beq
{\cal M}_d {\cal M}^\dag_d \sim
\left(\begin{array}{ccc} 1&1&1\\ 1&1&1\\ 1&1&1\\ \end{array}\right).
\eeq
For the leading approximation, the $V_R$ is
\beq
V_R \sim \left(\begin{array}{ccc}
 0 & -\frac{1}{\sqrt{2}} &  \frac{1}{\sqrt{2}}\\
 -\frac{\sqrt{2}}{\sqrt{3}} & \frac{1}{\sqrt{6}} &  \frac{1}{\sqrt{6}}\\
 \frac{1}{\sqrt{3}} & \frac{1}{\sqrt{3}} &  \frac{1}{\sqrt{3}}\\
\end{array} \right)
\eeq
and it is equivalent to all its  variants with any column permutation
or arbitrary row sign flips because
\beq
{\cal M}_d {\cal M}^\dag_d = {\cal S}_i {\cal M}_d {\cal M}^\dag_d
{\cal S}_i,\hs
({\cal M}^{dig}_d)^2 =  {\cal G}_i  ({\cal M}^{dig}_d)^2 {\cal G}_i
\eeq
where
\beqa
{\cal S}_1=\left(\begin{array}{ccc} 0&0&1\\ 0&1&0\\ 1&0&0\\ \end{array}
\right),
{\cal S}_2=\left(\begin{array}{ccc} 0&1&0\\ 1&0&0\\ 0&0&1\\ \end{array}
\right),
{\cal S}_3=\left(\begin{array}{ccc} 1&0&0\\ 0&0&1\\ 0&1&0\\ \end{array} \right)
\eeqa
correspond to the interchanging of $V_R$'s  1-3, 1-2, and 2-3 columns.
And ${\cal G}_1=diag(-1,1,1)$, ${\cal G}_2=diag(1,-1,1)$ and ${\cal G}_3=diag(1,1,-1)$
are to flip sign of first, second and third row of $V_R$.
This discrete symmetry is broken when  one allows for small corrections to ${\cal M}_d$
as discussed before.
Sometimes the first two indices switch to get the right order of
charged lepton mass, $m_e<m_\mu$.

From our discussions above, we see there is a 5D non-SUSY orbifold $SU(5)$
model setup which can give a  good charged fermion mass
hierarchy and mixing angles. Furthermore, this framework preserves the advantages
of $SU(5)$ GUT and avoid the major obstacles in the traditional 4D theories.
Since it's not SUSY, the unification of coupling constants is nontrivial.
We will defer the discussion of RG running and gauge unification to section 4.
Now we turn to the neutrino masses problem.

\section{Neutrino Masses}
It is well known that the symmetric $\mathbf{15}$ Higgs can
be used to generate neutrino Majorana masses
and break $(B-L)$ in conventional $SU(5)$ theories
\cite{Barbieri:1979ag}.
For the orbifold version, we  closely follow the study of
\cite{Chang:2003sx} where a  model of neutrino masses in 5D
orbifold $SU(3)_W$ unified theory without right handed singlet was constructed
and the symmetrical  $\mathbf{6}$ scalar filed was responsible for the
neutrino masses.
In the present case because $\mathbf{5}\times \mathbf{5}=\mathbf{10}+\mathbf{15}$,
besides the $\mathbf{15}$ Higgs, the anti-symmetrical $\mathbf{10}$ Higgs can also
work for generating the Majorana masses through one loop diagram,
see  Fig.\ref{fig:Zeeloop}.

\FIGURE[ht]{
\label{fig:Zeeloop}
\epsfxsize=3 in
\centerline{\epsfbox{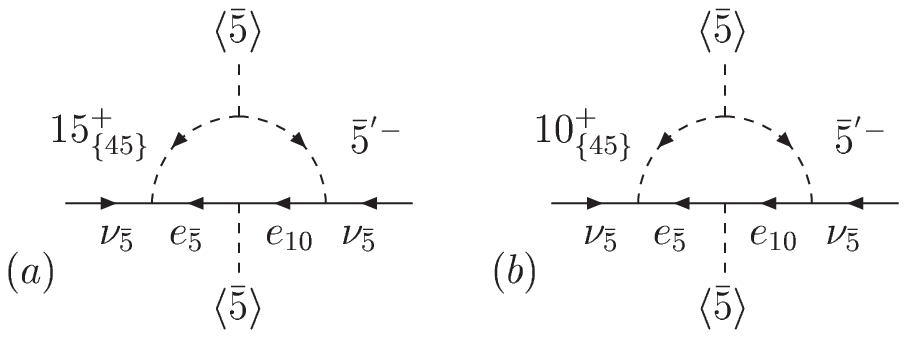}}
\caption{The 1-loop neutrino mass through
(a)$H_5^{'\dag} H_{15} H_5^*$ and (b)$H_5^{'\dag} H_{10} H_5^*$ coupling.
}}

Two things differ from  the $SU(3)_w$ case studied by
\cite{Chang:2003sx}. Firstly, in the $SU(3)_w$ theory both
$\mathbf{3}$ and $\mathbf{6}$ contribute one Higgs doublet.
One of the linear combination is the would be Goldstone boson
to be eaten by the SM gauge bosons and the orthogonal linear combination is the
required physical charged scalar that appears in the loop.
In $SU(5)$ case , the $\mathbf{10}$ and $\mathbf{15}$ break into
\beq
\label{eq:15chain}
\mathbf{15}=\underbrace{(6,1,-2/3)}_{P_{15}}+\underbrace{(3,2,1/6)}_{C_{15}}
 +\underbrace{(1,3,1)}_{T}
\eeq
and
\beq
\label{eq:10chain}
\mathbf{10}=\underbrace{(\bar{3},1,-2/3)}_{P_{10}}+\underbrace{(3,2,1/6)}_{C_{10}}+\underbrace{(1,1,1)}_{S}.
\eeq
in terms of SM quantum numbers.
For notational  simplicity, the symbols $P, C, S, T$ have been introduced to
indicate the specific component of  $\mathbf{10}$ or $\mathbf{15}$
as shown in Eqs.(\ref{eq:15chain},\ref{eq:15chain}).
We see that neither $\mathbf{10}$ nor $\mathbf{15}$ contain a Higgs doublet
component. So in the  $SU(5)$ case  we have to introduce another
$\mathbf{5'}$ Higgs for the radiative mechanism to work.

Secondly, due to  the $SU(5)$ symmetry
which imposes strong constraints on model building the parity of
exotic Higgs sector needs careful examination.
Now we discuss the various possible parity assignments of $\mathbf{15}$ and  $\mathbf{10}$.

We first assign the
$\mathbf{15}$ parity to be $(+-)$.
The   $\mathbf{15}(+-)$ will decompose into $P_{15}(+-)
+ C_{15}(++)+ T(+-)$.
By doing so, the triplet component  $T$ has no zero mode and
hence no VEV can be developed. Hence, there is no tree level neutrino Majorana mass.
Also, the hierarchy of the VEVs of SM Higgs doublet
and the $T$ Higgs triplet required by the
$\rho$ parameter measurement will be naturally avoided.
In this case, the $C_{15}$ color Higgs will improve
the unification of non-SUSY $SU(5)$ \cite{Murayama:1991ah}.

There are several shortcomings of this parity assignment. The Higgs boson running in the loop
(see Fig.\ref{fig:Zeeloop}) are both KK modes due to KK number conservation. So
the resulting neutrino masses are too small due to the suppression of KK mass $1/R$.
Furthermore  the parity of the extra $5'$ Higgs has to be $(+-)$
such that the bulk triple Higgs interaction
\beq
\label{eq:3H}
{m \over \sqrt{2M^*}} H^{'\dag}_5  H_{10/15} H^*_5 + H.c.
\eeq
is $Z'_2$ invariant.
Due to  the above Higgs potential proton decay now will be induced via the mixing of
the zero modes of $C_{15}$ and the color component $(3,1,-1/3)$ in $\mathbf{5}'$,
see Fig.\ref{fig:protonH}.
\FIGURE[ht]{
\epsfxsize=2in
\centerline{\epsfbox{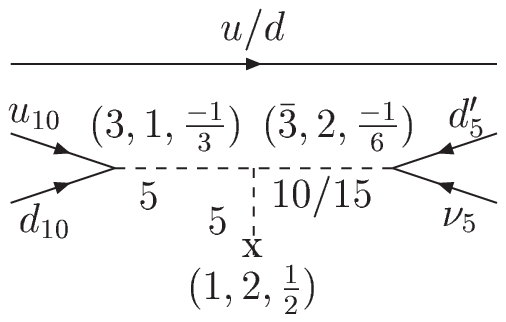}}
\label{fig:protonH}
\caption{
The proton and neutron decays through the mixing of $(3,1,-1/3)$
color Higgs in $\mathbf{5'}$ and the $(3,2,1/6)$ color Higgs boson
in $\mathbf{10}$ or $\mathbf{15}$.
}}

To satisfy both the constraints of proton decay and
the requirement that the resulting 1-loop neutrino mass
is of oder $.01$ eV, the zero mode mass of $C_{15}$
need to be as high as $10^{11}$ GeV and the triple Higgs coupling strength
$m$ is about $10^{16}$ GeV. This scenario is disfavored by the
strong interacting scalar sector and a very heavy scalar.

Next,  the parity of $\mathbf{15}$ can be chosen as $(-+)$.
Then the decomposition is $P_{15}(-+) + C_{15}(--)+T(-+)$.
Again, the triplet $T$ has no zero mode, so naturally avoid the VEV
problem and the tree level neutrino mass.
Now parity of the extra $5'$ Higgs has to be $(-+)$ correspondingly.
But the lepton number breaking component vanish at $y=0$
brane, so it doesn't couple to the $\mathbf{\bar{5}}$ living there.
For this scenario to work,  the $\mathbf{\bar{5}}$ fermions have to be placed at
$y=\pi R/2$ and enjoy no $SU(5)$ symmetry.
Unfortunately, even doing so, the resulting neutrino mass pattern is not right.

The third  parity assignment is $(--)$.
This leaves the $T$ components totally vanish at both branes.
It does not  work in our scenario where the $F_i$ are localized on the brane.

So we are only left with the $(++)$ assignment.
Explicitly, it satisfies the following transformation properties
\beqa
P H_{15} P^{-1}= + H_{15},\hs P' H_{15} P^{'-1} =+ H_{15}.
\eeqa

Now there are $P^{-2/3}_{15}(++)$, $(C^{+2/3}_{15},C^{-1/3}_{15})(+-)$
and  $(T^{+2},T^{+1},T^0)^T(++)$ Higgs.
The $T^0$ couples two neutrinos which could  give them the tree-level Majorana mass.
To avoid that, we assume that it is a regular scalar and does not develop a VEV.
Also, the zero mode of $P_{15}$ must be heavy to suppress the $K-\bar{K}$ mixing,
which can be gleamed from Fig.\ref{fig:Kmass}.
This gives a satisfactory neutrino mass matrix and other phenomenological constraints.

Most of above discussions on the parity  assignments can be applied to $\mathbf{10}$
so we will not repeat the analysis here but note that $\mathbf{10}$ doesn't acquire VEV.
The suitable $\mathbf{10}$ is of $(++)$ parity under $Z_2\times Z'_2$ too.
It has components $P^{-2/3}_{10}(++)$, $(C^{+2/3}_{10},C^{-1/3}_{10})(+-)$
and $S^{+1}(++)$.

Let's summarize the setup of the scalar sector:
(1) One extra $\mathbf{5'}$ Higgs.
(2) Either $\mathbf{15}$ or $\mathbf{10}$ scalar with parity
$(++)$ under $(Z_2\times Z'_2)$.
(3) The scalar $\mathbf{15}$ must not have VEV to avoid excessive fine tuning.
 The zero modes of scalar $\mathbf{15}$ or $\mathbf{10}$ must be heavy to
 avoid large contribution to  $K-\bar{K}$ mixing.

The necessary lepton number violation  terms arise from  the brane Yukawa interactions
\beqa
{\cal L}_{Y15}= \delta\left(y\right)\left[
{ \tilde{f}^{15}_{ij}\over \sqrt{2M^*}}\overline{\psi^{\{A\}c}_{\bar{5}i}}\psi^{\{B\}}_{\bar{5}j}
H^{\{AB\}}_{15} + H.c.\right]
\eeqa
for $\mathbf{15}$ where $H^{\{AB\}}_{15}=H^{\{BA\}}_{15}$ and
\beqa
{\cal L}_{Y10}= \delta\left(y\right)\left[
{ \tilde{f}^{10}_{ij}\over \sqrt{2M^*}}\overline{\psi^{\{A\}c}_{\bar{5}i}}\psi^{\{B\}}_{\bar{5}j}
H^{\{AB\}}_{10} + H.c.\right]
\eeqa
for $\mathbf{10}$ and
$H^{\{AB\}}_{10}= - H^{\{BA\}}_{10}$.
After integrating out the extra dimension, we then  parameterize  the relevant
4D effective interaction as:
\beq
{\cal L} \supset  {\tilde{f}^{10}_{ij}\over \sqrt{\pi R M^*} }\overline{e^{'c}}_i \nu'_j S^{+1}
 + { \tilde{f}^{15}_{ij}\over \sqrt{\pi R M^*} }\overline{e^{'c}}_i \nu'_j T^{+1} +  H.c.
\eeq
in the leptons' flavor  basis.
It is easy to see that  $\tilde{f}^{15}_{ij}=\tilde{f}^{15}_{ji}$ and
$\tilde{f}^{10}_{ij}= -\tilde{f}^{10}_{ji}$.
The entries of the  neutrino Majorana mass matrix
can be approximately calculated by ignoring the lepton masses in the propagators and are
given as follows:
\beq
\label{eq:rawmass}
({\cal M})^\nu_{ij}
= {1\over 16\pi^2}{ m(v_b)^{3/2} \over 2\sqrt{\pi R M^*}\sqrt{M^*}}
\sum_k { m_k  f_{ik}^{'5} \tilde{f}_{jk}^{10/15}\over M_1^2-M_2^2  }\ln\frac{M_2^2}{M_1^2}
\eeq
where $\tilde{f}^{10/15}$ is the 5D Yukawa coupling of either $\mathbf{10}$ or $\mathbf{15}$,
$m_k$ the mass of charged lepton-$k$ and
$M_1,M_2$ the masses of Higgs running in the loop.
And $f^{'5}$ is the effective 4D Yukawa coupling for the relevant
physical charged Higgs interaction:
\beq
f^{'5}_{ij} \bar{e}'_{R j}\nu'_i H^- + H.c.
\eeq

It is more  convenient to express the neutrino mass matrix in the
basis of charged lepton's mass eigenstates.
Alone the line of  the discussions in Sec.2, we expect the lepton Yukawa coupling
$(f^{'5})^T$ to exhibit the same pattern as $y_d$ in Eq.(\ref{eq:masspattern-D}).
As we rotating the charged leptons into their mass eigenstate,
the Yukawa matrix $f^{'5}$ is more or less  diagonalized in the same time
and it is proportional to the mass of charged lepton:
\beq
y^{'5}_{diag,ij}\sim \delta_{ij} {g_2 m_i \over \sqrt{2} M_W}.
\eeq
In the lepton mass eigenbasis, the Yukawa couplings for $\mathbf{10}$ or
$\mathbf{15}$ can be obtained by the following transformations
\beq
\label{eq:diagF}
f^{10} = V_R^\dag \left\{\tilde{f}^{10}\right\} V_R^*,\hs
f^{15} = V_R^\dag\left\{ \tilde{f}^{15}\right\} V_R^*.
\eeq
Only $V_R$ is involved since we have invoked the $SU(5)$ mass relation between
the $d^c$ and $e$.
Note that the antisymmetry of flavor indices $i,j$ of $\tilde{f}^{10}_{ij}$ and
the symmetricalness of  $\tilde{f}^{15}_{ij}$ is not affected by these transformations.

Therefore the Eq.(\ref{eq:rawmass}) is further simplified to
\beq
\label{eq:NuMatrix}
({\cal M})^\nu_{ij}
\sim {g_2\over 64\pi^2 M_W}{ m v_0 \over (\pi R M^*)}
 { m_i^2 f_{ij}^{10/15}\over M_1^2-M_2^2
 }\ln\frac{M_2^2}{M_1^2}+(i\leftrightarrow j)
\eeq
Thus, the matrix to diagonalize Eq.(\ref{eq:NuMatrix}) is the MNS matrix and
we can directly use the phenomenological results of neutrino oscillation studies.
With $v_0\sim 250$ GeV, $m_i\sim m_\tau$, $f^{10/15}\sim 1$, $M_1\sim 10^5$ GeV,
 $M_2\sim 200$ GeV and $(\pi R M^*)\sim 100$,  the overall neutrino mass can be
estimated to be $0.13 \times (f^{10/15})(m/\mbox{TeV})$ eV.
It can easily  give neutrino overall mass $\sim 0.01$ eV by adjusting
$m\sim 0.1$ TeV such that the triple scalars is weakly interacting, see Eq.(\ref{eq:3H}).

For the case  of $\mathbf{15}$, the neutrino mass matrix takes the form:
\beq
{\cal M}_\nu^{15}\sim\frac{1}{M}
\left( \begin{array}{ccc}
f^{15}_{11} m_e^2 &  f^{15}_{12}( m_e^2+m_\mu^2)/2 & f^{15}_{13}(
m_e^2+m_\tau^2)/2\\
f^{15}_{12}( m_e^2+m_\mu^2)/2 & f^{15}_{22}m_\mu^2 &f^{15}_{23}(
m_\mu^2+m_\tau^2)/2\\
f^{15}_{13}( m_e^2+m_\tau^2)/2 & f^{15}_{23}(m_\mu^2+m_\tau^2)/2 & f^{15}_{33}m_\tau^2\\
  \end{array}\right)
\label{eq:NuMass15}
\eeq
where $M$ is a mass to be determined by Eq.(\ref{eq:NuMatrix}).
Basically, it can  yield any of the acceptable neutrino mass matrices.
Here, we give few examples to illustrate the richness of resulting
neutrino mass pattern by using $\mathbf{15}$.

For instance, one solution is
\beq
\tilde{f}^{15} = \left( \begin{array}{ccc}
0.4959& 0.2012&-0.7524\\
0.2012& 0.2717 & -0.4316\\
-0.7524& -0.4316& 1.2360\\
  \end{array}\right)
\eeq
which leads to a  normal hierarchy mass pattern:
\beq
\overline{m}_0^{-1}{\cal M}_\nu = \left( \begin{array}{ccc}
3\times10^{-5}& 0.4233&-0.0271\\
0.4233& 1.0 & 0.9319\\
-0.0271& 0.9319& 1.1493\\
  \end{array}\right)
\eeq
with $\theta_1=37.4^\circ$, $\theta_2=44.04^\circ$,
$s_3=-0.13$, $\tri_\odot/\tri_{atm}=0.017$.
$\overline{m}_0$ sets the overall  neutrino mass scale which is determined by factors in
Eq.(\ref{eq:NuMatrix}).
In this example, $\overline{m}_0 = 0.05(m/ 0.15 \mbox{TeV} )$ eV.
Notice that the first entry will predict an unobservable rate for
neutrinoless double beta decay.
We have searched numerically for larger values of $({\cal M}_{\nu}^{15})_{11}$
and found solutions with first entry as big as $\sim 0.1$.
But they require fine tuning of Yukawa of order $10^{-4}$.
Even this value is below the sensitivity of the next generation of these experiments.

For neutrino mass matrices of the inverted hierarchy type, this
mechanism for generating neutrino masses requires fine tuning of Yukawa couplings
$f^{'15}_{ij}$ which we will discuss below.
Following  the classification in \cite{He, Alta03}, we explore the following two
leading patterns which lead to inverted hierarchy:
\beq
{\cal M}^{B1}_\nu
\sim {\overline{m}_0 \over \sqrt{2}} \times \left( \begin{array}{ccc}
  0&1&1\\1&0&0\\ 1&0&0\\
\end{array}\right),\hs
{\cal M}^{B2}_\nu
\sim \overline{m}_0 \times \left( \begin{array}{ccc}
  1&0&0\\0&\frac12&\frac12\\0&\frac12&\frac12\\
\end{array}\right).
\eeq
Both of the leading patterns give bi-maximal mixing angles
which conflict with the current experiments.
Some small entries in the structure zeros as well as small perturbations
to the leading terms are  necessary to accommodate the experimental data.

Let's examine two examples:
For $B1$ type, one solution for  ${\cal M}_\nu$ is
\beq
{\sqrt{2}\over \overline{m}_0}{\cal M}_\nu= \left( \begin{array}{ccc}
0.42& 1 & 0.922\\
1&0.097 & -0.464\\
0.922& -0.464& 0.006\\
  \end{array}\right)
\eeq
which gives $\theta_1=36.57^\circ$, $\theta_2=42.43^\circ$, $s_3=0.06$
and $\tri_\odot/\tri_{atm}=0.021$ and implies
\beq
\tilde{f}^{15}_{B1} = \left( \begin{array}{ccc}
3\times10^{-6}& 7\times 10^{-5} & -5\times 10^{-5}\\
 7\times 10^{-5}& 0.6667 &-0.4715\\
- 5\times 10^{-5}& -0.4715& 0.3335\\
  \end{array}\right).
\eeq

One  example of $B2$ type mass matrix is:
\beq
{{\cal M}_\nu \over \overline{m}_0}= \left( \begin{array}{ccc}
1& 0.02 & -0.01\\
0.02&0.49 & 0.5\\
-0.01& 0.5& 0.5\\
  \end{array}\right)
\label{eq:fb1}
\eeq
which gives $\theta_1=34.36^\circ$, $\theta_2=45.29^\circ$,
$s_3=0.021$ and $\tri_\odot/\tri_{atm}=0.030$.
The Yukawa coupling in the lepton weak basis is
\beq
\tilde{f}^{15}_{B2} = \left( \begin{array}{ccc}
6\times10^{-6}& -3\times 10^{-6} & -6\times 10^{-6}\\
 -3\times 10^{-6}& 0.6667 &-0.4714\\
- 6\times 10^{-6}& -0.4714& 0.3333\\
  \end{array}\right).
\label{eq:fb2}
\eeq

From the above examples we see that for both cases elements of the  first
row and first column are much smaller than the rest.
This amounts to fine tuning of the model.
We note in passing both inverted hierarchy types can be obtained
from  a Yukawa matrix with the leading structure that looks like

\beq
f^{15}\sim \left(\begin{array}{ccc}
1&\times  &\times\\
\times &\times &\times\\
\times &\times &\times \\
\end{array} \right)
\eeq
where $\times$ denotes a small number of order $10^{-5}$.

For $\mathbf{10}$, the resulting neutrino mass matrix structure is:
\beq
{\cal M}_\nu^{10}\sim \frac{1}{M}
\left( \begin{array}{ccc}
0 &  f^{10}_{ 12}( m_\mu^2-m_e^2)/2 & f^{10}_{13}(m_\tau^2-m_e^2)/2\\
f^{10}_{12}(m_\mu^2-m_e^2)/2 & 0 & f^{10}_{23}(m_\tau^2-m_\mu^2)/2\\
f^{10}_{13}(m_\tau^2-m_e^2)/2 & f^{10}_{23}(m_\tau^2-m_\mu^2)/2 & 0\\
  \end{array}\right).
\eeq
Again, $M$ is some mass to be determined by Eq.(\ref{eq:NuMatrix}).
The diagonal zeros are the result of antisymmetry
of the $\mathbf {10}$ representation. It can only give $B1$ type mass pattern, namely of the inverted
hierarchy type, if the Yukawa exhibit the following hierarchy:
$f^{10}_{d12}:f^{10}_{d13}:f^{10}_{d23}\sim 1:m_\mu^2/m_\tau^2: m_e^2/m_\tau^2$
 which implies that
\beq
\tilde{f}^{10} \sim \left( \begin{array}{ccc}
0& -\frac{1}{\sqrt{3}} & \frac{1}{\sqrt{6}} \\
\frac{1}{\sqrt{3}}& 0&  -\frac{1}{\sqrt{2}}\\
-\frac{1}{\sqrt{6}} & \frac{1}{\sqrt{2}}& 0 \\
  \end{array}\right).
\eeq
The hierarchy is also a reflection of the approximate $L_{\tau}+L_{\mu}-L_e$ symmetry.
It's impossible to get realistic neutrino masses by $\mathbf{10}$ alone at this
leading order.
However,  small diagonal entries can be generated through 2-loop
correction which  we shall leave  for future works.
Another  obvious mean is to include $\mathbf{15}$ as the perturbation
source but this will lead to non-minimal models.

Hence, it is interesting that $\mathbf {15}$ gives a normal hierarchy
whereas the $\mathbf {10}$ give inverted hierarchy of neutrino masses without
additional symmetry or fine tuning of their respective Yukawa couplings.

\section{Gauge Unification }
It is well known that the minimal 4D $SU(5)$ unification is ruled out by current data.
Even the supersymmetric version \cite{susy-gut, five} is disfavored \cite{Mura-P}.
With the extended Higgs sector required to generate neutrino masses we expect the
unification of the gauge couplings will be a challenge.
Since our model is a 5D one albeit non-supersymmetric, the renormalization group
(RG) running of the SM gauge couplings crosses several thresholds and scales.
We follow here the methods developed by the works of
\cite{Hall:2001pg,Dienes:1998vg}.
Below the compactification scale, $1/R$, we have 4D effective field  theory with
thresholds crossings from the $\mathbf{15}$ or $\mathbf{10}$ mandated by our model of
neutrino masses.
This will constitute an intermediate scale between that of electroweak
breaking and $1/R$.
Between $1/R$ and the unification scale one has power law running of the 5D gauge theory.
Clearly the problem of gauge unification now becomes highly nontrivial and
requires careful treatment which we detail below.
We will only discuss RG running and unification up to one loop.

The first task is to establish the lower limit on  the mass of
the  $\mathbf{10}$ or the $\mathbf{15}$, generically called $M_P$.
We found the most stringent limit for $M_P$ comes from the $K_L-K_S$ mass
difference mediated by exchanging color Higgs components of
$P_{10,15}$, see Fig.\ref{fig:Kmass}.

\FIGURE[th]{
\epsfxsize=3in
\centerline{\epsfbox{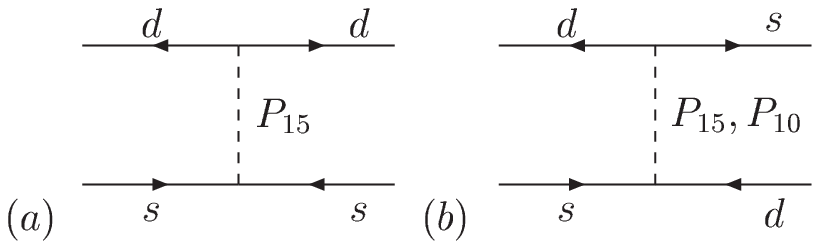}}
\label{fig:Kmass}
\caption{The tree level contribution to Kaon mass difference
by  exchanging the $(6,1,-2/3)$ Higgs in $\mathbf{15}$
representation ((a),(b) ) or the $(\bar{3},1,-2/3)$ Higgs in
$\mathbf{10}$ representation ( (b) ). }}

The $\tri S \neq 0$ relevant terms are:
\beq
{\tilde{f}^{10}_{ij}\over \sqrt{\pi R M^*}}\bar{d}^{\alpha}_i
d^{\mathbf{c}\beta}_{j}P_{10}^{[\alpha\beta]}
+{\tilde{f}^{15}_{ij}\over \sqrt{\pi R M^*}}\bar{d}^{\alpha}_i
d^{\mathbf{c}\beta}_{j}P_{15}^{\{\alpha\beta\}}+H.c.
\eeq
where $\alpha,\beta$ are the color indices and it is understood the above
expression shall be made $SU(3)$ invariant (for  details see \cite{splitf}).
The Yukawa interaction can be rotated into down quarks' mass
eigenstates again by  Eq.(\ref{eq:diagF}) because the $SU(5)$
symmetry relates left handed leptons and right handed down quarks.
After applying  the Fierz transformation, we arrive at the effective $\tri S=2$
operator from exchanging the $P$ component of $\mathbf{15}$:
\beq
{\cal L}^{\tri S=2}\sim \frac12 \frac{1}{M_{P}^2}
\left[ {f^{15}_{11} f^{15\dag}_{22}+ |f^{15}_{12}|^2 \over \pi R M^*}\right]
(\bar{d}_R^\alpha\gamma_\mu s_R^\alpha)(\bar{d}_R^\beta\gamma^\mu s_R^\beta)+H.c.
\label{eq:dS=2}
\eeq
By using vacuum insertion approximation, the resulting
kaon mass difference can be estimated  to  be:
\beq
\tri M_K\sim \frac23 f_K^2 m_K Re
\left( { f^{15}_{11} f^{15\dag}_{22}+|f^{15}_{12}|^2 \over 2M_P^2\pi R M^{*}} \right).
\eeq
The formula applies to $\mathbf{10}$ except that $f^{10}_{ij}$
has no diagonal element.
Plugging in the experimental values: $f_K\sim 0.16$ GeV, $ m_K=
0.4976$ GeV and $\tri M_k =3.48\times 10^{-15}$ GeV
we obtain the lower limit for $M_P$:
\beq
M_P>1.11\times 10^{6}
\left( {f^{15}_{11} f^{15\dag}_{22}+ |f^{15}_{12}|^2 \over \pi R M^* }\right)^{1/2}
\left( {\tri M_K^P\over 3.48\times 10^{-15}\mbox{GeV}}\right)^{-1/2}\mbox{GeV}.
\eeq
where $\tri M_K^P$ is the mass difference due to color scalar contribution.
Similar analysis can be naively extended to $\tri M_B$ by substituting
$f_K \leftrightarrow f_B$ and $M_K \leftrightarrow M_B$ which gives a less  constrained
limit:
\beq
M_P>4.33\times 10^{5}
\left( {f^{15}_{11} f^{15\dag}_{33}+ |f^{15}_{13}|^2 \over \pi R M^* }\right)^{1/2}
\left( {\tri M_B^P\over 3.75\times 10^{-13}\mbox{GeV}}\right)^{-1/2}\mbox{GeV}.
\eeq
Evidently, the limit goes down as the absolute value of specific Yukawa couplings of
$\mathbf{15}$ or $\mathbf{10}$ is lower.

If we take the above number as the intermediate scale the question arises whether these
color Higgs will induce rapid proton decay.
It can be checked that there are neither tree level nor one loop contributions.
Hence, with a relatively low value of  $M_P \sim 10^{5}$ GeV, obtained by assuming that
$\pi R M^*\sim 100$, $\tri M_K$ is saturated by the contribution of exotic scalars
and the extreme case $|f^{10/15}|\sim 1$, will not run afoul of proton stability.

The one loop  gauge coupling RG running after passing
various thresholds can be written as
\beqa
\label{eq:RGeq}
\alpha_i^{-1}(\mu) = \alpha_i^{-1}(M_Z)
- {a^{SM}_i \over 2\pi}\ln{\mu\over M_Z}
- {a^{H}_i \over 2\pi}\ln{\mu\over M_H}\nonr\\
- {\tilde{a}^o_i \over 2\pi}\sum_{n=1}^{N^o}\ln{\mu R\over 2n-1}
- {\tilde{a}^e_i \over 2\pi}\sum _{n=1}^{N^e} \ln{\mu R \over 2n }
\eeqa
The last two $\tilde{a}$ terms are  the KK modes contributions
when energy scale crosses $1/R$.
The integers $N^o$ and $N^e$ are the highest odd( $(+-)$ and $(-+)$ )
and even( $(++)$ and $(--)$ ) KK excitation level
below the scale $\mu$
\beq
{2N^e \over R}\leq \mu,\hs
{2N^o-1 \over R}\leq \mu.
\eeq

For the SM, the beta functions are  well known:
$a^{SM}_i=(4, -10/3, -7)+ n_H\times (1/10,1/6,0)$ where
$n_H$ is the number of Higgs doublet zero modes.
At scale $M_Z$, deriving from $\alpha(M_Z)^{-1}=127.934(7)$ and
$\sin^2\theta_W=0.231113(15)$\cite{PDG}, we have
\beq
\alpha_1^{-1}=\frac35{ \cos^2\theta_W\over\alpha}=59.031(35),\hs
\alpha_2^{-1}={\sin^2\theta_W\over\alpha }=29.568(17)
\eeq
and  $\alpha_s(^-1)=8.53$.
In our model, the zero modes of $\mathbf{10}$ or $\mathbf{15}$ Higgs has a large mass
$M_{P}$ imposed by phenomenology.
Below $M_{P}$, only SM particles go into the beta function.
As one passes  the threshold, $M_{P}<\mu<\frac{1}{R}$, the zero modes of
$\mathbf{10}$ or $\mathbf{15}$ will  contribute to the beta function.
For $\mathbf{15}$, it has zero modes $P_{15}(6,1,-2/3)$ and $T(1,3,1)$ so
\beq
(a_1^{15}, a_2^{15}, a_3^{15})=\left(\frac{17}{15}, \frac23, \frac56\right).
\eeq
For $\mathbf{10}$, the zero modes are $P_{10}(\bar{3},1,-2/3)$ and
$S(1,1,1)$ hence
\beq
(a_1^{10}, a_2^{10}, a_3^{10})=\left(\frac{7}{15}, 0, \frac16\right).
\eeq
Finally, when the scale is over  the compactification scale, various KK excitations
gradually come in and contribute to RG running.
For very high KK levels, summing over the KK tower roughly gives the power
law running of coupling constants.
This power law behavior can be easily understood by adding up the KK
excitations level by level.
By simple algebra and  Stirling's approximation, the KK sums are
\beqa
S^o&=&\sum_{n=1}^{N^o}\ln{\mu R\over 2n-1}= N^o \ln(\mu R) - \ln{(2N^o-1)! \over
(N^o-1)!} + (N^o-1)\ln 2\nonr\\
&\sim& N^o \ln(\mu R)-N^o \ln 2N^o + N^o -\ln\sqrt{2} +{\cal
O}(\frac{1}{N^o})\\
S^e&=&\sum_{n=1}^{N^e}\ln{\mu R\over 2n}=N^e \ln(\mu R) -\ln N^e!
-N^e\ln2\nonr\\
&\sim& N^e\ln(\mu R)-N^e\ln 2N^e + N^e -\ln\sqrt{2\pi N^e}+{\cal
O}(\frac{1}{N^e}).
\eeqa
When $\mu \gg 1/R$, $N^o\sim N^e \sim \mu R/2$, the contribution of all the
KK excitations to $\alpha_i^{-1}(\mu)$ can be  approximately calculated as follows:
\beq
 -{\tilde{a}_i \over 4\pi }(\mu R -\ln 2) +{\tilde{a}^e_i \over
 4\pi}\ln\left({\pi \mu R \over 2}\right)
\eeq
where $\tilde{a}_i=\tilde{a}^e_i+\tilde{a}^o_i$ is the sum of the
beta function from even and odd KK components.
As $\mu$ being higher then the compactification scale the full $SU(5)$
symmetry start to emerge.
So it is not surprising  that any complete $SU(5)$ multiplet gives equal
amount to three gauge running,
$\tilde{a}_1=\tilde{a}_2=\tilde{a}_3$, which will not affect  the
gauge coupling unification.
As unification is concerned, only the even (or equivalently the odd )
KK components matter.
Explicitly, we list the beta functions of all even KK
contents in Table 2.

\TABULAR[ht]{c c l c c c}{
  Field & $\tilde{a}$& Even Components &$\tilde{a}^e_1$ & $\tilde{a}^e_2$ & $\tilde{a}^e_3$ \\
\hline\hline
$H_5$ & $1/6$ & $H_w(1,2,1/2 )$& $1/10 $ & $1/6$ & $0$ \\
\hline
$H_{15}$ & $7/6$ & $P_{15}^{-2/3}(6,1,-2/3)$ & $8/15$ &0  & $5/6$ \\
 & & $T(1,3,1)$ & $3/5$ & $2/3$ & $0$ \\
\hline
$H_{10}$ & $1/2$ & $P_{10}^{-2/3}(\bar{3},1,-2/3)$ & $4/15$ &0  & $1/6$ \\
 & & $S(1,1,1)$ & $1/5$ & $0$ & $0$ \\
\hline
 && $G^\mu(8,1,0)$ & $0$ & $0$ & $-11$ \\
&& $W^\mu(1,3,0)$ & $0$  & $-22/3$ & $0$ \\
$SU(5)$ Gauge& $-35/2$& $A^\mu(1,1,0)$ & $0$ & $0$ & $0$ \\
&&$X^5,Y^5(3,2,-5/6)$ & $5/12$ & $1/4$ & $1/6$ \\
&&$X^5,Y^5(\bar{3},2,5/6)$ & $5/12$ & $1/4$ & $1/6$ \\
\hline
$T_{bulk}$ &2& $U^c(\bar{3},1,-2/3)$ & $16/15$ & $0$ & $2/3$ \\
&&$E^c(1,1,1)$ & $4/5$ & $0$ & $0$ \\
\hline
$T'_{bulk}$ &2& $Q'_L(3,2,1/6)$ & $2/15$ & $2$ & $4/3$ \\
\hline}{RG contribution from various KK excitations.}

Note that the fifth  components of the KK $X,Y$ gauge fields come in as
real scalars which give one half of the contribution of complex scalars.
Also note that the combined result of bulk filed $T$ and $T'$ are same to
three coupling running so it will not affect unification.
The contributions of KK modes  can be summarized as:
\beqa
(\tilde{a}^e_1, \tilde{a}^e_2, \tilde{a}^e_3 )=
\left(\frac56,-\frac{41}{6},-\frac{32}{3}\right)
+ n_5\times \left(\frac{1}{10},\frac16,0\right)\nonr\\
+ n_{15}\times \left(\frac{17}{15},\frac23, \frac56\right)
+ n_{10}\times \left(\frac{7}{15},0, \frac16 \right)
\eeqa
and
\beq
\tilde{a}=-\frac{35}{2}+\frac16 n_5 +\frac76 n_{15}
+\frac12 n_{10} +4 n_T
\eeq
where $n_5$, $n_{10}$, $n_{15}$  are the numbers of bulk $\mathbf{5, 10, 15}$ Higgs
and  $n_T$ is the number of generation of ten-plet fermion in bulks respectively.
With  the minimum particle content we cannot obtain  gauge unification as can be
seen explicitly in  Fig.5.

To make this model work for unification, we propose to use vector fermions.
The vector nature is to ensure no new anomaly is induced.
They can be either $n_Q$ pairs of heavy quark doublets, $Q_H+ \overline{Q}_H$,
coming from  of $T_H(+-)+\overline{T}_H(+-)$
or $n_L$ pairs of heavy lepton doublets $L_H+\overline{L}_H$
from bulk $n_L$ pairs of $F_H(++)+\overline{F}_H(++)$.
The new $F_H+\overline{F}_H$ must be engineered  not to  mix with the
$F_{1,2,3}$,  otherwise the neutrino mass matrix will be spoiled.
For this reason we favored heavy quark pairs.

For simplicity, we assume they all have a common zero mode mass,
denoted as $M_{QL}$,  which can be adjusted to achieve unification.
When $M_{QL}<\mu<1/R$, the threshold effect of $Q_H$s
\beq
(\tri a_1,\tri a_2,\tri a_3 )=  2 n_Q\times \left( {1\over 15}, 1,\frac23 \right)
\eeq
should be considered and the extra term
\beq
\tri \alpha_i^{-1}(\mu) = - {\tri a_i \over 2\pi}\ln{\mu \over M_{QL}}
\eeq
should be  included in Eq.(\ref{eq:RGeq}).
As the scale larger then $1/R$, their KK modes give extra
contribution to $\tilde{a}^e$ and $\tilde{a}^o$ as
\beq
\tri \tilde{a}^e_i= 2 \tri a_i,\hs
(\tri \tilde{a}^o_1, \tri \tilde{a}^o_2, \tri \tilde{a}^o_3 )=
4 n_Q\times\left( \frac{14}{15},0,\frac13 \right)
\eeq
where another factor $2$ is due to that 5D fermions are vector like.

It turns out that a simple solution is with $n_Q=1$.
Now we have two $\mathbf{5}$ Higgs, one $\mathbf{15}$ Higgs
with  mass $M_P=10^{5}$ GeV to evade the constraint from $\tri M_K$
and one pair of $Q_H+\overline{Q}_H$.
If we require that $(\pi R M_{GUT})=100$, then the solution can be found
to be $M_{QL}=18.03$ TeV and $M_{GUT}=8.94\times 10^{15}$GeV, see Fig.6.

Note the splitting of even and odd KK contribution slows down the
convergence and push the unification scale to higher end.
It is interesting to note that  the volume factor, $\sim (\pi R M_{GUT})$,
is related to $M_{QL}$.
A  higher $M_{QL}$ provides smaller volume factor and lower $M_{GUT}$.
For example, to have $(\pi R M_{GUT})=1000$ it requires $M_{QL}=6.0$ TeV
and $M_{GUT}=2.97\times 10^{16}$ GeV.
We have the approximate upper bound  $M_{QL}<117.1$ TeV and
$M_{GUT}> 1.16\times 10^{15}$ GeV from the consistency  requirement that $\mu > 1/R$.

\DOUBLEFIGURE[ht]
{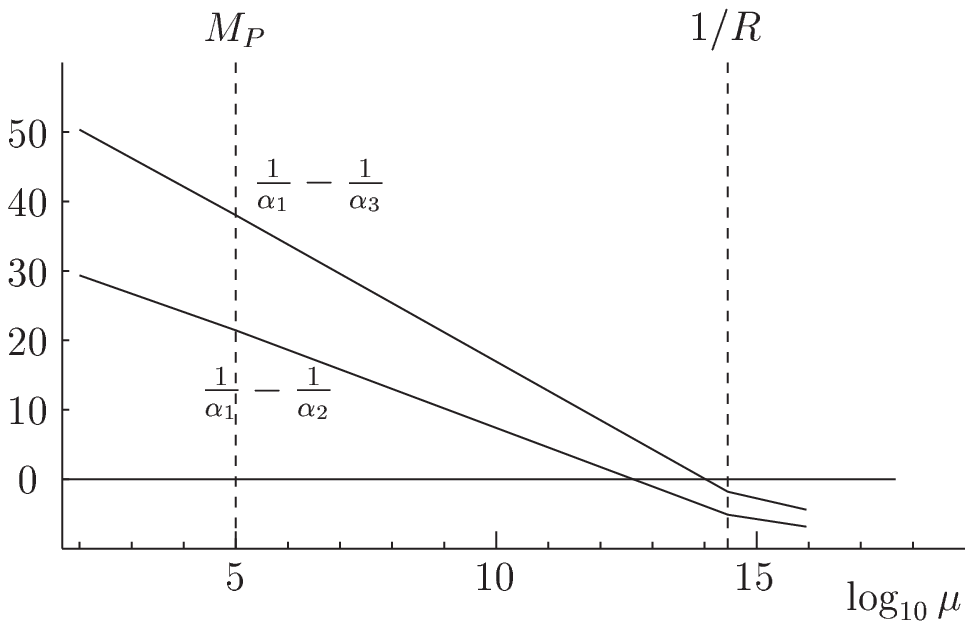, width=2.5in}
{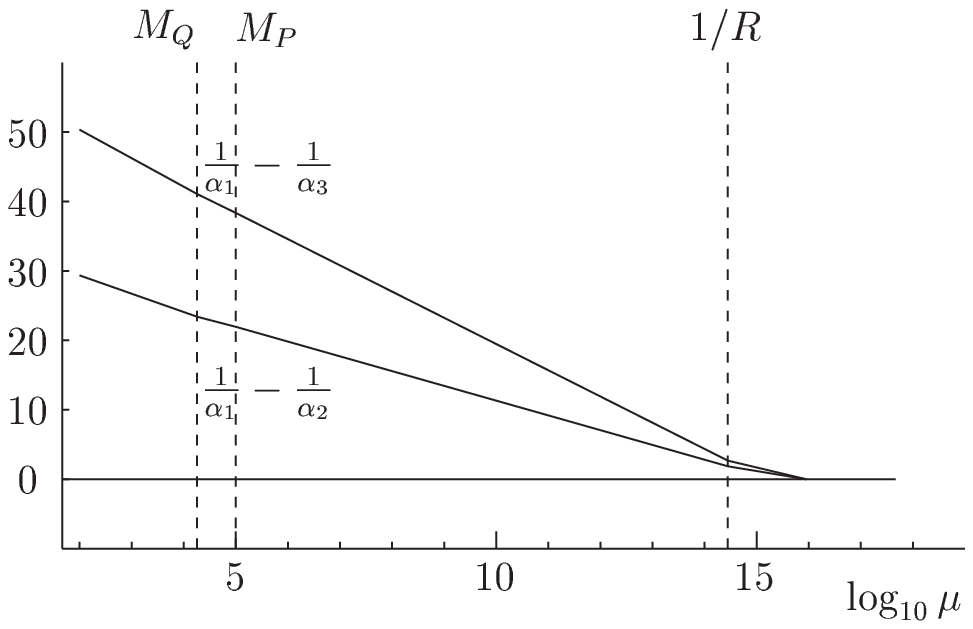, width=2.5in }
{The RG running with $M_{P}=10^5$ GeV  and $1/R=2.81\times 10^{14}$ GeV.}
{The RG running with $M_{P}=10^5$ GeV,
$M_Q=18.03$ TeV  and $1/R=2.81\times 10^{14}$ GeV.
They converge  at $8.94\times 10^{15}$GeV.}

We note in passing that unification can still be made by adding
many bulk $\mathbf{5}$ Higgs with parity $(++)$ if heavy fermion doublets are not used.
However, this solution requires a large  number($\geq 6$) of extra scalars needed.
We deem this to be an unpalatable solution.

\section{Phenomenology}
\FIGURE[ht]{
\epsfxsize=1 in
\centerline{\epsfbox{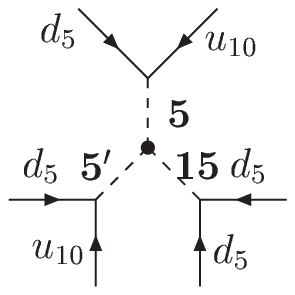}}
\label{fig:NNbar}
\caption{The Feynman diagrams for $N\hskip-1mm-\hskip-1mm \bar{N}$
oscillation.}}

The neutron anti-neutron oscillation may be induced by the mixing of
Higgs, see Fig. \ref{fig:NNbar}.
The relevant 6-quark operator can be expressed as
\beq
G_{\tri B=2} u^c u^c d^c d^c d^c d^c +H.c.
\eeq
the parameter $G_{\tri B=2}$ is of mass dimension minus five.
It can be estimated to be:
\beq
G_{\tri B=2} \sim {m f_5^2  f_{15} \over (1/R)^4 M_P^2(\pi R M^*)}<
{ 10^{-52}\over (\pi R M^*)}\mbox{GeV}^{-5}
\eeq
in arriving the result, the mass $M_P\sim10^{5}$ GeV from the constraint
of $\tri M_K$, $f_5\sim f_{15}\sim 1$, and $m\sim 1/R\sim 10^{14}$ GeV
have  been plugged in.
It is safely within  the experimental limit\cite{Baldo-Ceolin:1994jz}
\beq
\tau_{N-\bar{N}}> 0.86 \times 10^8 sec
\eeq
or equivalent, $G_{\tri B=2}< 3\times 10^{-28} \mbox{GeV}^{-5}$.

\FIGURE[ht]{
\epsfxsize=3in
\centerline{\epsfbox{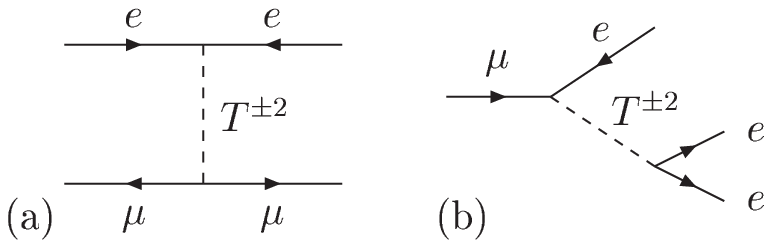}}
\label{fig:mu3e}
\caption{The Feynman diagrams for (a)muonium-antimuonium  and (b) $\mu\ra3e$
 transition by exchange Higgs $T^{\pm2}$.}}

The lepton flavor violating muonium ($\mu^+e^-\equiv M$)-antimuonium
($\mu^-e^+\equiv \overline{M}$) transition and rare decay $\mu\ra 3e$
can be induced at tree level by exchanging the $T^{\pm2}$ scalars which
belong to $\mathbf{15}$ Higgs, see Fig.\ref{fig:mu3e}.
The interaction of $T$ with charged leptons is given by
\beq
{\cal L}_T= {f^{15}_{ij}\over \sqrt{\pi R M^*}} \bar{l^c}_i l_j T^{+2} + H.c.
\eeq
where Yukawa coupling $f^{15}$ is in the charged lepton's mass basis.
Assuming there is no external electromagnetic fields and the
mass difference of the mixed state, $\delta$ , is small.
The possibility of  observing a  the transformation of muonium
into antimuonium can be written as
$P(\overline{M})\sim \delta^2 / 2 \Gamma_\mu^2 $
where $\Gamma_\mu$ is the muon decay rate.
And the mixing can be estimated to be
\beq
\delta\equiv 2 \langle \overline{M}|{\cal H}_{M\overline{M}}|M \rangle
={2 f^{15}_{11}f^{15\dag}_{22}  \over \pi a^3 M_T^2(\pi R M^*)}
\eeq
where $a$ is the Bohr radius ($a^{-1}=\alpha m_e$).
$P(\overline{M})$ can be expressed as
\beq
P(\overline{M})\sim 1.1\times 10^{-16}
{|f^{15}_{11}f^{15\dag}_{22}|^2 \over (\pi R M^*)^2 }\left({M_T \over 10^5\mbox{GeV}}\right)^{-4}
\eeq
which is safely within the current experimental limit
$P(\overline{M})<8.3\times 10^{-11}$\cite{MAMT}.

The $\mu\ra 3e$ transitions can be described by
an effective lagrangian
\beq
 {(f^{15\dag}_{11} f^{15}_{12})\over M_T^2 (\pi R M^*)} (\bar{e^c} \mu_L)(\bar{e}e^c
 )+ H.c.
\eeq
which leads to  the branching ratio of $\mu\ra 3e$:
\beq
Br(\mu\ra 3e)
= {2|f^{15\dag}_{11} f^{15}_{12}|^2\over  g_2^4(\pi R M^*)^2}
\left( {M_W \over M_T}\right)^4.
\eeq
The above equation can be expressed in term of $\tri M_K$ to eliminate
the uncertainty of the absolute value of $f^{15}$:
\beqa
Br(\mu\ra 3e)= 2\left({3 M_W^2 \tri m_K \over g_2^2 f_K^2 m_K}\right)^2
\times \left({\tri m^P_K \over \tri m_K}\right)^2
\times \left(\frac{M_P}{M_T}\right)^4\nonr\\
\times \left|{|f^{15\dag}_{11}f^{15}_{12}| \over
f^{15\dag}_{11}f^{15}_{22}+|f^{15}_{12}|^2}\right|^2
\eeqa
where $\tri m^P_K$ is the contribution to kaon mass difference by exchanging $P$ scalar.
We give explicit dependence of the ratio of $M_P$ to $M_T$ because we expect
they will split after quantum corrections are taken into account.
In the above we have used the relations between Yukawa couplings and the elements
of the neutrino mass matrix, see Eq.(\ref{eq:NuMass15}).
So the ratio of Yukawa couplings can be replaced by the ratio of the
corresponding elements in $\cal {M}_{\nu}$:
\beq
Br(\mu\ra 3e)\sim 3.02\times 10^{-16} \left({\tri m^P_K \over \tri m_K}\right)^2
\left(\frac{M_P}{M_T}\right)^4
\left({2 m_{11}m_{12} \over m_{11}m_{22} + (2
\frac{m_e}{m_\mu}m_{12})^2}\right)^2.
\eeq
It's straightforward to extend the analysis  to  $\tau\ra 3l$ transitions.
Assuming that the hierarchy of the elements of neutrino mass matrix is
smaller than factor 100, this model predicts
\beqa
Br(\mu\ra3e):Br(\tau\ra 3e):Br(\tau\ra3\mu):Br(\tau\ra\mu
e e):Br(\tau\ra e\mu\mu)\nonr\\
\sim \frac{m_{12}^2}{m_{22}^2}:
\left(\frac{m_\mu}{m_\tau}\right)^4\frac{m_{13}^2}{m_{22}^2}:
\left(\frac{m_e}{m_\tau}\right)^4\frac{m_{23}^2}{m_{11}^2}:
\left(\frac{m_\mu}{m_\tau}\right)^4\frac{m_{23}^2}{m_{22}^2}:
\left(\frac{m_e}{m_\mu}\right)^4\frac{m_{12}^2}{m_{11}m_{22}}.
\eeqa
The extra suppression of mass ratio to the fourth power makes it
very difficult to find experimental signal in $\tau\ra 3l$ decays.
Hence, $\mu\ra 3e$ is the best probe of the model.
But this model exhibits an interesting characteristic: only the neutrino mass
matrices  of $B1$ type have the chance to be benefited from large enhancement
$(m_{12}/m_{22})^2$ such that $\mu\ra 3e$ can be observed in near future experiments.

The rare decays $\mu\ra e\gamma$, $\tau\ra\mu\gamma$, $b\ra s\gamma$ etc can be
induced by one loop diagrams which involve corresponding components
of either $\mathbf{15}$ or $\mathbf{10}$.
Clearly,  these rare decays are useful tools for probing  flavor physics.
But, again, these process are suppressed by  $(M_W/M_P)^4$ plus the loop factor
suppression which make their rates too small to be tested in  foreseeable experiments.

What about seeing the effects of the heavy quarks $Q_H+\overline{Q}_H$? Since
we expect their masses to be heavier then 10 TeV direct production of these
particles is not likely at the LHC. One can look for their virual effects.
Because they are vector like so they have no leading order
contribution to the $S$ and $T$ parameters\cite{Peskin:1991sw}.

\section{Conclusions}
We have explicitly constructed  viable models of neutrino mass matrix involving only three
active SM Weyl neutrinos  without introduction of singlet
right-handed fermion in 5D orbifold $SU(5)$ unification models.
The crucial source of lepton number violation is the $\mathbf {15}$ or $\mathbf {10}$
bulk Higgs fields.
This  model preserves the orbifold solution to the triplet-doublet problem
and avoids rapid proton decay mediated by leptoquarks at the tree level due to
KK number conservation at the vertex coupling two  $\mathbf{10}$ fermions which are
bulk fields. However, at the one loop level proton decay can occur.
Our estimate puts it safely within the experimental bound.

In the class of models we studied the overall scale of neutrino mass is partly
controlled  by the Yukawa couplings of the exotic Higgs bosons to the brane leptons
and triple scalar coupling and the mass scale of the exotic Higgs fields.
It is of order $10^{-2}$ eV after all the phenomenological constraints are satisfied.
We found that the $\mathbf {15}$ prefers a normal hierarchy; whereas the
$\mathbf {10}$ gives an inverted hierarchy when  two loop effects are taken into account.
Otherwise only bi-maximal mixing can be obtained. Since the model is restricted to
three Weyl neutrinos the mass matrix obtained is necessary Majorana. This is a necessary
but not sufficient condition to induce neutrinoless double beta
decay of nuclei. To do so the $(11)$ entry of the mass matrix must be non vanishing.
Generically without fine tuning of the Yukawa couplings, it is typically of order
$10^{-4}$ eV which puts it outside the range of detectability in the next round of experiments.
This is a reflection of the fact that normal hierarchy or Zee-like mass matrices are
preferred. Stating this differently, if a value of $|m_{ee}|\sim 0.1$ eV is extracted
from the next generation of experiment which we imagine to give positive signatures
then we would require a 4 orders of magnitude hierarchy in the Yukawa couplings
of $\mathbf{15}$ for this mechanism to work.

As expected the introduction of exotic Higgs exacerbates the problem of gauge
unification of $SU(5)$.
We found a  solution to this in the 5D model by using a pair of vector bulk
fermions $\mathbf{10}$ and $\overline{\mathbf{10}}$.
The masses of these fermions are in the 10 TeV range and unification occurs
at $\sim 10^{16}$ GeV and the compactification radius $1/R \sim 10^{14}$ GeV.
 On the other hand, these exotic scalars are promising source to give
the universe enough matter anti-matter asymmetry. Detailed analysis of this will
appear elsewhere \cite{changA}.

In our study we have not considered in detail the origin of the charged fermion
mass hierarchy.
It is sufficient for us to put the third family on the $SU(5)$ brane and the first
two families  $\mathbf {10}$ fermions as bulk fields.
Perhaps this can solved by the split fermion scenario similar to
\cite{splitf} but this we leave for future considerations.

It is well known challenge  to test the physics of
various models for neutrino masses. Currently the data do not distinguish
the different classes of models let alone the many within one framework.
For the seesaw model, generally the right-handed Majorana neutrinos are of order
of GUT scale and direct probe is out of the question.
For the case of bulk right-handed neutrinos studies thus far done have indicated
phenomenological tests are only possible if the compactification is very
large \cite{Dienes:1999,bulkph}.
Since the bulk neutrino behaves like a sterile ones more structures in the oscillation
pattern will be a good signature \cite{bulknu}.
Currently no such structure is seen and more precise measurements will be needed.
 In our radiative scenario it is possible  to see the effects of lepton number
violating Higgs in rare decay.
We found that once the mass  $M_P$ satisfies  the constraint from $\tri M_K$
then  most of the lepton and baryon  flavor  violating transition
induced by $\mathbf{15}$ are far below the present experimental limits. This
also puts it out of the range of direct detection in the near future.
However, the decay $\mu\ra 3e$ can be just below the current experimental limit.
If found may indicate the neutrino mass matrix is of $B1$ type inverted
hierarchy which leads to a very low neutrinoless double beta decay rate.

\acknowledgments
Work was supported in part by the Natural Science and Engineering Council of
Canada.
WFC  wishes  to thank  the National Center for Theoretical Science in Taiwan
for its hospitality where part of this work was done.

\newpage


\end{document}